\documentclass{iaus}
\usepackage{graphicx}
\def\sch{Schwarzschild}

\title[Gamma Ray Bursts] 
{Gamma Ray Bursts: basic facts and ideas}

\author[Gabriele Ghisellini]   
{Gabriele Ghisellini
}
\affiliation{Osservatorio Astronomico di Brera, Via Bianchi 46 Merate I--23807 Italy
\\ email: {\tt gabriele.ghisellini@brera.inaf.it} }

\pubyear{2010}
\volume{275}  
\pagerange{1--8}
\jname{Jets at all Scales}
\editors{G.E. Romero, R.A. Sunyaev \& T. Belloni}
\begin{document}

\maketitle

\begin{abstract}
The recent years witnessed a dramatic improvement in our knowledge 
of the phenomenology and physics of Gamma Ray Bursts (GRBs).
However, our ``pillars of knowledge" remain a few, while 
many aspects remain obscure and not understood.
There is no general agreement on the radiation mechanism
of the prompt emission, nor on the process able to convert
the bulk motion of the fireball into random energy of the
emitting leptons.
The afterglow phase can now be studied at very early phases,
showing an unforeseen phenomenology, still to be understood.
In this context, the detection of $\sim$GeV emission from $\sim$10\% of
GRBs, made possible by the {\it Fermi} satellite, can hopefully  
shed light on some controversial issues.
\keywords{gamma rays: bursts, radiation mechanisms: nonthermal, supernovae: general}
\end{abstract}

\firstsection 


\section{Pillars of knowledge}

What are the fundamental and not controversial facts
characterizing Gamma Ray Bursts?
I propose a list of seven ``pillars" of knowledge, selected
in an admittedly completely subjective way, following
this criterion:
If we did not know this particular fact, would we lose a 
basic piece of knowledge?

\subsection{GRBs are cosmological}

One of the major achievements of the {\it Beppo}SAX satellite
was to localize a GRB with enough accuracy to make the 
pointing of an optical telescope possible, allowing to find the redshift.
At the same time, the afterglow was discovered (Costa et al. 1997,
for GRB 970228).
As we know, the first measured redshift was $z=0.835$ for GRB 970508
(Metzeger et al. 1997; the redshift for GRB 970228 was measured later,
due to the faintness of its host galaxy).

This ended a long and animated  discussion about the origin of GRBs
(i.e. ``local" , i.e. associated to neutron stars in the Galactic halo,
or cosmological, as predicted by Paczynski 1986), and finally 
set the power of these objects:
they are indeed the most explosive events of the Universe after the Big Bang.
Soft $\gamma$--ray repeaters, instead, were found to be ``nearby" magnetars
undergoing flares, and associated to supernova remnants.

One of the early successes of the {\it Swift} satellite was to localize
short GRBs, and therefore allow the optical follow up leading to establish that
they, also, are cosmological events (Gehrels et al. 2005).

\begin{figure}[h]
\vspace*{-1.0 cm}
\hspace*{-0.8 cm}
\includegraphics[width=6in, height=4in]{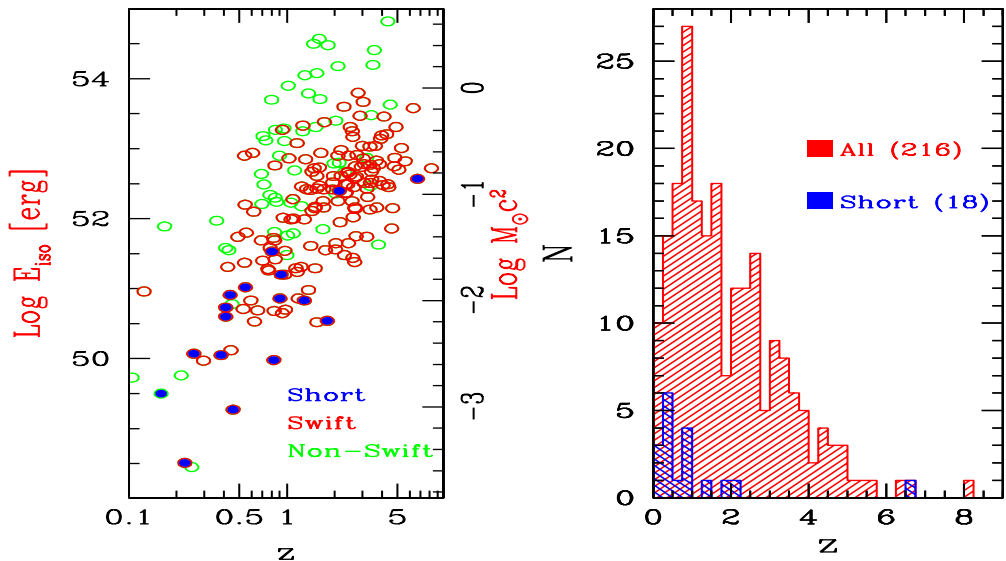} 
\vspace*{-1.8 cm}
\caption{
Left: Energetics of the prompt emission of GRBs with measured redshift.
Be aware that the plotted energetics are simply the observed fluences
multiplied by $4\pi d_{\rm L}^2/(1+z)$, so they are only lower limits
to the bolometric $E_{\rm iso}$.
Right: Redshift distribution.
}
\label{fig1}
\end{figure}

The top panel of
Fig. \ref{fig1} reports the energetics of the GRBs with measured redshifts,
and the bottom panel shows the redshift distribution for long and short GRBs.
Most of them have been detected by {\it Swift}.
With the caveat that the shown isotropic energetics $E_{\rm iso}$
are not bolometric ones,
nor have been K--corrected, we can see that the largest $E_{\rm iso}$ 
correspond to more than a solar mass entirely converted into energy.
Short GRBs with measured $z$ are still very few, but they seem to lie
closer and to be less energetic than long ones.

\subsection{GRBs have large bulk Lorentz factors}

GRBs are the fastest extended objects of Nature, with bulk Lorentz factors 
$\Gamma$ that can exceed 1000.
The first evidence came from theory: injecting a colossal amount of energy
in a small volume (of the order of a few \sch\ radii of size) leads
inevitably to the formation of electron--positron pairs that makes
the so--called ``fireball" opaque to the huge internal pressure.
The fireball is then obliged to expand, becoming relativistic with
$\Gamma\propto R$ until the internal energy is converted into bulk motion
(if the fireball remains opaque).

A nice observational evidence of relativistic speeds came, in the
late '90s, from the behavior of the radio afterglow of a few GRBs,
whose light curve varied wildly for $\sim$3 weeks, ``calming down"
after this time.
This behavior was immediately interpreted as due to radio scintillation
quenched by the increasing size of the radio source.
Therefore it was possible to establish the expansion velocity,
that turned out to be superluminal, requiring $\Gamma>4$ after 3 weeks 
from the trigger (Frail et al. 1997).

A very recent evidence came instead from the detection of GRBs in the GeV
energy range by the LAT instrument onboard the {\it Fermi} satellite.
The GeV flux partly overlaps with the emission at lower energies
detected by the other {\it Fermi} instrument 
(the GBM, sensitive in the 8 keV--30 MeV range).
If the two emissions are cospatial, then the variability shown in the GBM dictates
that the source must have a minimum $\Gamma$--factor, to avoid
$\gamma$--$\gamma \to e^\pm$ suppression of high energy photons.
For these LAT--detected GRBs (10\% of the total), the minimum
$\Gamma$--values are around 1000. (Abdo et al. 2009; Ackermann et al. 2010).

If, instead, the GeV emission is not cospatial with the GBM one,
then it is very likely that it belongs to the afterglow phase.
The short delay between the GBM and LAT emission can be due to the
required time for the onset of the afterglow (i.e. the deceleration time
of the fireball moving in the circumburst medium).
The shorter this time, the higher the $\Gamma$--factor.
Again, for all the LAT detected bursts, values around 1000 are derived
(Ghirlanda, Ghisellini \& Nava 2010; Ghisellini et al. 2010).

\subsection{Prompt plus afterglow emission phases}

The GRB emission has two phases, the erratic, $\gamma$--ray 
(or hard X--ray) {\it prompt phase}, and a smoother {\it afterglow} phase.
This means that not all the energy of the fireball is radiated away 
during the prompt, but some remains.
This was predicted before the first observations of the afterglow,
but probably at the same time of the detection of long duration GeV emission by the EGRET
instrument onboard the {\it Compton Gamma Ray Observatory} satellite 
(Meszaros, Rees \& Papathanassiuou 1994) 
and then elaborated by Meszaros \& Rees (1997); Vietri (1997); Sari \& Piran (1997).

The fact that there are two emission phases suggests that there must be two mechanisms
at work, one for the prompt and one for the afterglow.

\subsection{Long and short}

The duration of the prompt emission of GRBs is bimodal, with a minimum around
2 seconds.
In astrophysics bimodal distributions are always looked at with suspicion,
since malicious selection effects can be at work.
The convincing arguments of a real bimodality comes from the spectrum since
short GRBs are harder than long ones. 
This was first evident from the hardness ratio (i.e. the ratio of the flux in two energy bands;
Kouveliotu et al. 1993)
and then substantiated by direct spectral analysis (Ghirlanda, Ghisellini \& Celotti 2004).
The bimodality suggests that GRBs come in two flavors, in turn suggesting two 
different operating mechanisms, and possibly two kinds of progenitors.
The prevalent idea is that long GRBs originate immediately after the collapse
of a massive, Wolf--Rayet star, while short GRBs originate from the merging
of two compact objects.

\subsection{Spikes have same durations}

This ``pillar" is not very popular, but it was nevertheless crucial 
for the development of the current leading scenario of ``internal shocks" (see below)
explaining the prompt emission.
The evidence is that the light curves of GRBs (both long and short) often
shows spikes of emission, whose duration $\Delta t_{\rm spike}$ is on average the same
(Ramirez--Ruiz \& Fenimore 2000).
In other words, there is no lengthening of $\Delta t_{\rm spike}$ with $t$, the
time since the trigger.
Emission episodes, on average, should then involve regions of similar sizes,
and then probably at the same distance from the central engine.

\subsection{Supernova connection}

We believed that long GRBs are associated to Supernovae Ib,c,
but not all SN Ib,c are associated to GRBs (Soderberg et al. 2006
estimated a fraction less than 1\%).
The evidence comes from spectroscopy (for nearby events)
and re--brightening of the optical light curve (up to $z\sim$1).
The association strongly indicates that the progenitor of long GRBs
is a massive stars, that has lost its hydrogen and helium envelopes.
But there are at least two nearby bursts (GRB 060614, Gal--Yam et al. 2006, and  
GRB 060505, Ofek et al. 2007) where the SN was not found.
If present, it would be at least two orders of magnitude less luminous
than SN1998bw (associated to GRB 980425).

\subsection{Common behaviors and trends}

``When you see a GRB, you see just one GRB" was a popular motto
in the past, meaning that all GRBs were different, with no common
behaviors.
Now this is not true any longer, and there are indeed common trends and
similarities.
Just two examples: the spectral energy relation, linking $E_{\rm peak}$ to
the prompt energetics or peak luminosity, and the typical behavior of the
early (i.e. less than a day or so) X--ray afterglow, with its characteristic
``steep--flat--steep" light curve (Tagliaferri et al. 2005),
and superimposed on that, $\sim$1/3 of GRBs 
show X--ray flares (Burrows et al. 2007).
These similarities are the starting point for any serious and general modelling:
ideas are in fact abundant, but with no clear prevalence of one over the others.

\section{Ideas and enigmas}

\subsection{Central engine}
The prevalent idea is that long GRBs are caused by the collapse
of a Wolf--Rayet star leading to the formation of a black hole of a few solar masses
rapidly spinning.
This black hole accretes 0.1--1 $M_\odot$ from a dense surrounding torus
for a time more or less equal to the duration of the prompt emission.
There are several energy reservoirs: neutrinos, the gravitational
energy of the infalling matter, and the rotational energy of the newly formed
black hole.
The latter is the greatest, since it amounts to 
$\sim 0.29 M_{\rm BH} c^2\sim 5.3\times 10^{53} (M_{\rm BH}/M_\odot)$.
The problem is how to extract it efficiently.
The leading idea it to use the Blandford \& Znajek (1977) process,
for which a super--critical magnetic field of $B\sim 10^{15}$ G is required.
For short GRBs, the merging scenario assumes two compact objects (e.g. two neutron stars)
forming a $\sim 2 M_\odot$ black hole surrounded again by a dense accreting torus.
The central engine can then be the same for long and short bursts.

Although prevalent, this is not the only idea.
Instead of a black hole, one could have, at least initially, a neutron star,
(that collapses into a black hole only later, as a re--edition of the
Vietri \& Stella (1998) Supranova model).
This has been proposed both to explain precursors (Wang \& Meszaros 2007,
see Burlon et al. 2008 for the characterization of precursors).
A magnetar has been proposed to explain the flat (plateaux) 
phase of the early X--ray afterglow (e.g. Lyons et al. 2010).
An even more radical idea was put forward by Paczynski \& Haensel (2005),
who suggested a quark star as the central engine.
These authors pointed out that the surface of such a star acts as a one--way 
membrane, since baryons can only enter, but not escape.
Leptons and magnetic fields, instead, can escape.
This would help to explain the paucity of baryons in the fireball
(i.e. the baryon ``loading" problem).

\subsection{Magnetic or matter dominated?}

In the most popular scenario a huge amount of energy is injected
into a small volume. 
Due to the colossal internal energy (and the inevitable
creation of $e^\pm$ pairs, making the fireball opaque to radiation)
the fireball is bound to accelerate with $\Gamma\propto R$.
At the same time the comoving temperature ($T^\prime \propto 1/R$)
decreases, and when it goes below $\sim$20 keV almost all the pairs annihilate
without being re--created.
Still, a small amount of protons and their accompanying electrons ensures
that the fireball continues to be opaque until the internal energy is
entirely converted into bulk motion.
Thus we need another mechanism to re--convert the bulk energy into radiation.
This is provided by collisions of different parts of the outflowing relativistic
wind moving with different $\Gamma$--factors.
This are the so--called  ``internal shocks", occurring at $R\sim 10^{12}$--$10^{14}$ cm,
where the fireball has turned transparent (for Thomson scatterings).
Then we have a disorder $\to$ order $\to$ disorder  process.
Lyutikov \& Blandford (2003; see the review by Lyutikov 2006 and references therein)
advocated instead a simpler order $\to$ disorder process:
the acceleration is due to a dominating magnetic field, allowing for almost matter free
fireballs.
One clear test to distinguish is the so--called ``optical flash", occurring
when the fireball starts to be decelerated by the interstellar medium:
if it is matter dominated, then a reverse shock develops, that originates
an important, and fastly decreasing, emission component (predicted in the
optical or in the IR), that would be absent in magnetically dominated fireballs.
Indeed optical flashes have been seen, but in a very small fraction of bursts,
so the issue is unsettled.
Another diagnostic would be polarization of the prompt emission (see the review
by Lazzati 2006), that awaits for hard X--rays polarimeters and observations.

\subsection{Internal shocks?}

Collisions between different parts of the relativistic wind 
(Rees \& Meszaros 1994) leading to ``internal" dissipation
is the leading idea for the dissipation mechanism for the prompt emission.
What can be dissipated is only the {\it relative kinetic energy} of the
two colliding parts or shells.
The process has then a ``built in" low efficiency (e.g. Lazzati et al. 1999).
Consider also that, as a result of the dissipation, we distribute
the available energy to protons, magnetic fields and leptons.
Only the energy given to the latter can be efficiently transformed
into radiation.
After the different parts of the fireball have collided and ``merged",
the fireball runs into the circumburst medium, originating a forward ``external"
shock.
This collision is with not--moving material, and should be much
more efficient, since the entire kinetic energy of the fireball can be used.
The foreseen densities and energies ensure that leptons initially
radiatively cool rapidly (fast cooling regime).
Thus, at least initially, the resulting afterglow is an efficient radiator.
The energy radiated by the afterglow should then be greater than the energy
radiated during the prompt phase.
We observe just the opposite (e.g. Willingale et al. 2007): 
$E_{\rm prompt}/E_{\rm afterglow}\sim 10$.

\subsection{Radiation process of the prompt}

The popular choice is that it is synchrotron emission.
Shocks should indeed accelerate electrons to relativistic energies,
and amplify magnetic fields, making the synchrotron option a natural one.
On the other hand, we also require any emission process to be efficient,
making the electron to cool completely in a timescale much shorter than any
conceivable dynamical or integration time.
The spectrum produced by a cooling electron population cannot be harder than
$F(\nu)\propto \nu^{-1/2}$. 
The spectra of practically all bursts are harder than that, and a minority
are even flatter than $\nu^{1/3}$, the low frequency tail of the
spectrum produced by a non cooling electron population (Preece et al. 1998;
Ghisellini et al. 2000).
Continuous re--acceleration or heating of the electrons is not compatible with the
idea of internal shocks (in which each electron is energized only once),
and other possible ``way--outs" (adiabatic expansion, steep gradients of the magnetic field,
contribution from synchrotron self--Compton) face  severe problems as well.
Alternatives have been proposed 
[such as jitter radiation (Medvedev 2000);
quasi thermal Comptonization (Ghisellini \& Celotti 1999; Giannios 2008);
bulk Compton (Lazzati et al. 2000);
multicolor blackbody (Peer \& Ryde 2010); effects of Compton cooling in the Klein Nishina limit
(Daigne  et al. 2010)], but there is no prevalent idea yet.

\subsection{Spectral energy correlations}

The time integrated spectrum of the prompt, in $\nu F_\nu$, has a well defined
peak at $E_{\rm peak}$, that
correlates with the isotropic energy of the prompt $E_{\rm iso}$:
$E_{\rm peak}\propto E_{\rm iso}^{1/2}$ (Amati et al. 2002).
When it is possible to measure the jet opening angle of the jet,
it is possible to estimate the collimated energy $E_\gamma$,
that correlates more tightly with $E_{\rm peak}$
($E_{\rm peak}\propto E_\gamma^b$, with $b=0.7$ for an homogeneous
circumburst density and $b=1$ for a wind-like profile;
Ghirlanda et al. 2004; Nava et al. 2006).
This correlation is tight enough to allow to ``standardize" GBRs
for their use as standard candles to constrain the cosmological parameters.
The physical reality of these correlations has been hotly disputed (see 
Ghirlanda, these proceedings, and reference therein), because selection effects
could play a role.
On the other hand, the observations of the same correlations {\it within
single GRBs} (Firmani et al. 2009; Ghirlanda et al. 2010) proves that 
a physical robust mechanism, still to be understood, is responsible
for these spectral--flux (or spectral--energy) correlations.

\begin{figure}[h]
\begin{center}
\includegraphics[width=4.5in, height=4.2in]{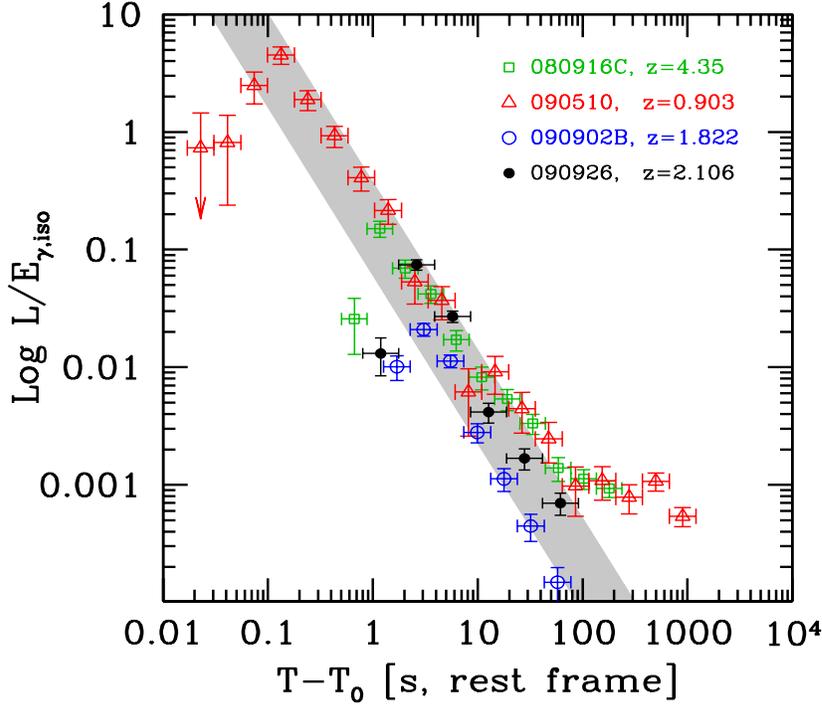} 
\vspace*{-0.3 cm}
\caption{The light curve of the four brightest bursts detected by the {\it Fermi}/LAT
at GeV energies. 
The luminosity has been divided by the energetics of the emission 
detected by the {\it Fermi}/GBM instrument ($\sim$MeV).
The time is in the rest frame of the sources. 
The grey stripe indicates a $t^{-10/7}$ slope.
The flattening at long times for GRB 090510 and 
GRB 080916C indicates the level of the background.
Note the similarity of the different light curves.
}
\label{fig2}
\end{center}
\end{figure}

\section{High energy emission}

EGRET, in the '90s, detected a handful of GRBs 
above 100 MeV, and since then we have been left with the question: 
does this emission belong to the prompt phase or is it afterglow
emission produced by the fireball colliding with the circum--burst
medium? Or has it still another origin?
A puzzling feature of the EGRET high energy emission
was that it was long lasting, yet it started during the prompt
phase as seen by BATSE.
{\it Fermi}/LAT is $\sim20$ times more sensitive, and indeed it detected
a dozen GRBs just in its first year of life.

\subsection{Common behaviors}

From the analysis of the first 12 GRBs detected by the LAT we have found these
properties (Ghisellini et al. 2010):

\vskip 0.1 cm
\noindent
{\bf Time delay --} Usually, the LAT emission lags the emission detected by the GBM
(from fractions of seconds, especially for short bursts, to a few seconds).

\vskip 0.1 cm
\noindent
{\bf Long lasting --} As already shown by the first EGRET detections, the 
emission seen by the LAT lasts for a longer time than the emission in the GBM.

\vskip 0.1 cm
\noindent
{\bf No spectral evolution --} 
The average, time integrated LAT photon index is close to 2,
with no evidence of strong spectral evolution.

\vskip 0.1 cm
\noindent
{\bf LAT and GBM spectral slopes are often different --}
The GBM data can be fitted with a Band function, composed
of two smoothly joining power laws. 
All but two bursts (GRB 080916C and GRB 090926)
have LAT slopes intermediate between the two slopes of the GBM fit.   

\vskip 0.1 cm
\noindent
{\bf LAT fluences are smaller than GBM ones --} 
The majority of bursts have LAT fluences
smaller than the GBM ones.
The two short bursts GRB 081024B and GRB 090510
and GRB 090902B have comparable LAT and GBM fluences.

\vskip 0.1 cm
\noindent
{\bf Common decay --}
Fig. \ref{fig2} shows the light curves of the 4 brightest GRBs with
redshift, once the 0.1--100 GeV luminosity is divided by 
the energetics $E_{\rm \gamma, iso}$ of the flux detected by the GBM.
The shaded stripe with slope $t^{-10/7}$ is shown for comparison.
These four GRBs are all consistent,
within the errors, with the same decay, both in slope and in normalisation.
Note that GRB 090510, a short burst, behaves similarly to the other 3 bursts,
that belong to the long class, but its light--curve begins much earlier.


\subsection{A radiative fireball?}
\vskip 0.2 cm
The above properties are just what expected by the afterglow emission due to
an external shock.
The short LAT--GBM delay can be caused by a large $\Gamma$ (close to 1000), making the fireball 
to decelerate at early observed times.
{\it The earlier the onset of the afterglow, the brighter the afterglow
at early times:} this explains why only 10\% of GRBs have been 
detected by the LAT: they correspond to bursts having the largest $\Gamma$--factors.

The relatively steep decay of the LAT light curves is very close to
what expected if the fireball is radiative (i.e. most of the dissipated energy
is radiated).
In this case the bolometric flux in fast cooling decays as
$F(t)\propto t^{-10/7}\sim t^{-1.43}$ (Sari et al. 1998; Ghisellini et al. 2010).
The slopes of the LAT spectra, being close to unity (in energy), 
are indeed a good proxy for the bolometric fluxes.

The large peak energy of the GBM flux suggests that electron--positron pairs 
might play a crucial role for the setting of the radiative regime:
a tiny fraction of the prompt photons, scattered (i.e. ``decollimated")
by the circumburst
electrons, are immediately converted into pairs by the high energy
photons of the prompt, largely increasing the lepton to proton ratio
(Beloborodov 2002).
When the shock comes, the dissipated energy can then be given mostly
to leptons rather than to protons, and this makes the fireball radiative.
Note a key ingredient of this scenario: to produce pairs efficiently, 
the prompt emission should have a sufficient number of photons above threshold,
i.e. above 511 keV.
 
\section{Conclusions}

The increase of knowledge can be represented as the volume
of an expanding sphere, so it goes like $R^3$, the cube of the radius. 
The surface of the sphere is the at the frontier with the unknown, the unexplained,
and usually goes like $R^2$.
But for GRBs it seems that that surface is a fractal, whose dimensionality
is greater than 2...
Is it good or bad? 
It may be perceived as depressing at first, after all these years of hectic studies,
but at a second sight it should be taken as a good opportunity, especially for the
younger scientists: there is still something really fundamental to be found out.

\begin{discussion}

\discuss{Mirabel}{What is the fraction of long GRBs with no Supernovae?}

\discuss{Ghisellini}{Difficult to say, since there is a clear observational
bias (only the nearby ones can be found).
I can answer the symmetric question: only $\sim$1\% of SN Ibc are associated to a GRB.} 

\discuss{Mirabel}{What is the range of masses for quark stars?}

\discuss{Ghisellini}{One is tempted to associate quark stars to the 2--5 solar mass range,
because the observed masses of neutron stars are below 2 solar masses, and the estimates
for Galactic black hole masses are larger than 5 solar masses. } 

\discuss{de Gouveia}{A comment and a question. The comment: you mention the work of Paczynski (2005)
proposing a quark star model to explain the engine of GRBs. In 2002 Lugones, Ghezzi, myself 
and Horvath published an ApJ Letter suggesting the same. 
I'm glad to see that the GRB community is starting considering this alternative possibility.}
The question: you have mentioned also the magnetar model for the engine but we know that there
are major theoretical constraints on the production of magnetars.
Could you comment on that?

\discuss{Ghisellini}{The Soft Gamma Ray Repeaters are rather convincingly associated
to magnetars. These systems can undergo major flares (once per century?) as the one
observed on December 27 2004 from SGR 1806--20.
If we put SGR 1806--20 at a few tens of Mpc, then we would classify its giant flare 
as a short GRB.
There has been discussion about what fraction of short GRBs are giant flares from
magnetars, but both spectral studies (the spectrum should be a blackbody) and
correlations with nearby galaxies suggested that this fraction must be small.
}

\end{discussion}

\end{document}